\documentclass[twocolumn,epjc3]
{svjour3}
\usepackage{braket}
\usepackage{amsfonts,amsmath,amssymb,bm,a4wide,graphicx,makeidx,multicol,color}

\smartqed  
\RequirePackage{graphicx}
\begin{document}

\title{Neutrino eigenstates and flavour, spin and spin-flavour oscillations in a constant magnetic field
}

\author{Artem Popov\thanksref{e1,addr1}
        \and
        Alexander Studenikin\thanksref{e2,addr1,addr2} 
}

\thankstext{e1}{e-mail: ar.popov@physics.msu.ru}
\thankstext{e2}{e-mail: studenik@srd.sinp.msu.ru}

\institute{Department of Theoretical Physics, \ Moscow State University,
        119991 Moscow, Russia \label{addr1}
           \and
           Joint Institute for Nuclear Research, 141980 Dubna, Russia \label{addr2}
}


\maketitle

\begin{abstract}
We further develop a recently proposed new approach to the description of the relativistic neutrino flavour $\nu_e^L \leftrightarrow \nu_{\mu}^L$, spin $\nu_e^L \leftrightarrow \nu_{e}^R$ and spin-flavour $\nu_e^L \leftrightarrow \nu_{\mu}^R$ oscillations in a constant magnetic field that is based on the use of the exact neutrino stationary states in the magnetic field. The neutrino flavour, spin and spin-flavour oscillations probabilities are calculated accounting for the whole set of possible conversions between four neutrino states. In general, the obtained expressions for the neutrino oscillations probabilities exhibit new inherent features in the oscillation patterns. It is shown, in particular, that: 1) in the presence of the transversal magnetic field for a given choice of parameters (the  energy and magnetic moments of neutrinos and the strength of the magnetic field)  the amplitude of the flavour oscillations $\nu_e^L \leftrightarrow \nu_{\mu}^L$ at the vacuum frequency is modulated by the magnetic field frequency, 2) the neutrino spin oscillation probability (without change of the neutrino flavour) exhibits the dependence on the mass square difference $\Delta m^2$.
It is shown that the discussed interplay of neutrino oscillations in magnetic
fields on different frequencies can have important consequences in astrophysical environments,
in particular in those peculiar for magnetars.
\end{abstract}

\section{Introduction}
\label{intro}
Massive neutrinos have nontrivial electromagnetic properties (see \cite{Giunti:2014ixa} for a review, the update can be found in \cite{Studenikin:2018vnp}). And for many years since \cite{Fujikawa:1980yx,Shrock:1982sc},  it is known that in the easiest generalization of the Standard Model the magnetic moment of the mass states of neutrinos is not zero \cite{Fujikawa:1980yx,Shrock:1982sc}:
\begin{equation}\label{mu_D}
    \mu^{D}_{ii}
  = \frac{3e G_F m_{i}}{8\sqrt {2} \pi ^2}\approx 3.2\times 10^{-19}
  \Big(\frac{m_i}{1 \ \mathrm{eV} }\Big) \mu_{B}.
\end{equation}
The best terrestrial upper bounds on the level of
$\mu_{\nu} < 2.9 \div 2.8 \times 10^{-11} \mu_{B}$
on neutrino magnetic moments are obtained by the GEMMA reactor neutrino experiment \cite{Beda:2012zz} and recently by the Borexino collaboration \cite{Borexino:2017fbd} from  solar neutrino fluxes. An order of magnitude more strict astrophysical bound on the neutrino magnetic moment is provided by the observed properties of globular cluster stars \cite{Raffelt:1990pj,Viaux:2013hca,Arceo-Diaz:2015pva}.

The neutrino magnetic moment precession in the transversal magnetic field ${\bf B}_{\perp}$ was first considered in \cite{Fujikawa:1980yx}  (this possibility was also mentioned in \cite{Cisneros:1970nq}), then the spin-flavor precession in vacuum was discussed in \cite{Schechter:1981hw}, the importance of the matter effect was emphasized in \cite{Okun:1986na}. The effect of the resonant amplification of neutrino spin oscillations in ${\bf B}_{\perp}$ in the presence of matter was proposed in \cite{Akhmedov:1988uk,Lim:1987tk}, the magnetic field critical strength the presence of which makes spin oscillations significant was introduced \cite{Likhachev:1990ki}, the impact of the longitudinal magnetic field ${\bf B}_{||}$ was discussed in \cite{Akhmedov:1988hd} and just recently in \cite{Fabbricatore:2016nec}. In a series of papers \cite{Pulido:1999xp,Akhmedov:2000fj,Akhmedov:2002ti,Akhmedov:2002mf} the solution of the solar neutrino problem was discussed on the basis of neutrino oscillations with a subdominant effect from the neutrino transition magnetic moments conversion in the
solar magnetic field (the spin-flavour precession).

Following to the general idea first implemented in \cite{Dmitriev:2015ega,Studenikin:2017mdh}, we further develop a new approach to the description of the relativistic neutrino flavour $\nu_e^L \leftrightarrow \nu_{\mu}^L$, spin $\nu_e^L \leftrightarrow \nu_{e}^R$ and spin-flavour $\nu_e^L \leftrightarrow \nu_{\mu}^R$ oscillations in the presence of an arbitrary constant magnetic field. Our approach is based on the use of the exact stationary states in the magnetic field for the classification of neutrino spin states, contrary to the customary approach when the neutrino helicity states are used for this purpose.

Within this customary approach the helicity operator is used for the classification of a neutrino spin states in a magnetic field. The helicity operator does not commute with the neutrino evolution Hamiltonian in an arbitrary constant magnetic field and the helicity states are not stationary in this case. This resembles situation of the flavour neutrino oscillations in the presence of matter when the neutrino mass states are also not stationary. In the presence of matter the neutrino flavour states are considered as superpositions of stationary states in matter. These stationary states are characterized by ``masses" ${\widetilde{m}}_{i}(n_{eff})$ that are dependent on the matter density  $n_{eff}$ and the effective neutrino mixing angle $\tilde{\theta}_{eff}$ is also a function of the matter density.

The proposed alternative approach to the problem of neutrino oscillations in a magnetic field  is based on the use of the exact solutions of the corresponding Dirac equation for a massive neutrino wave function in the presence of a magnetic field that stipulates the description of the neutrino spin states with the corresponding spin operator that commutes with the neutrino dynamic Hamiltonian in the magnetic field. In what follows, we also account for the complete set of conversions between four neutrino states.

\section{Massive neutrino in a magnetic field}
\label{sec:1}
Consider two flavour neutrinos with two helicities accounting for mixing
\begin{eqnarray}\label{nu_nu}
\nu_e^{L(R)} &=& \nu_1^{L(R)} \cos\theta + \nu_2^{L(R)} \sin\theta,\nonumber \\
\nu_{\mu}^{L(R)} &=& -\nu_1^{L(R)} \sin\theta + \nu_2^{L(R)} \cos\theta,
\end{eqnarray}
 where $\nu_i^{L(R)}$ are the helicity neutrino mass states, $i=1,2$. Recall that for the relativistic neutrinos the helicity states approximately coincide with the chiral states $\nu_i^{L(R)} \approx \nu_i^{ch^-(ch^+)}$.  As it is well known, the relativistic neutrinos produced in a weak process  are  almost in the left-handed helicity states. The detailed discussion on neutrino helicity and chirality can be found in \cite{Thomson2013}.
However, the helicity mass states $\nu_i^{L(R)}$ are not stationary states in the presence of a magnetic field. In our further evaluations we shall expand $\nu_i^{L(R)}$ over the neutrino stationary states $\nu_i^{-(+)}$ in the presence of a magnetic field.

The wave function $\nu_i ^s$ ($s=\pm 1$) of a massive neutrino  that propagates  along $\bf{n}_{z}$ direction in the presence of a constant and homogeneous arbitrary orientated magnetic field  can be found as  the solution of the Dirac equation
\begin{equation}\label{eq1}
  (\gamma_{\mu} p^{\mu} - m_i-{\mu_i}{\bm{\Sigma}\bm{B}})\nu_i^s (p)=0,
\end{equation}
where $\mu_i$ is the neutrino magnetic moment and the magnetic field
is given by ${\bf B}=(B_\bot,0,B_\|)$. In the discussed two-neutrino case the possibility for a nonzero neutrino transition moment $\mu_{ij} \ ( i\neq j)$ is not considered and two equations for two neutrinos states $\nu_i ^s$ are decoupled.
The equation (\ref{eq1}) can be re-written in the equivalent form
\begin{equation}
\hat{H_i}\nu_i ^s= E \nu_i ^s,
\end{equation}
where the Hamiltonian is
\begin{equation}\label{Ham}
\hat{H_i} = \gamma_0 \bm{\gamma}\bm{p} + \mu_i \gamma_0 \bm{\Sigma}\bm{B} + m_i \gamma_0.
\end{equation}
The spin operator that commutes with the Hamiltonian (\ref{Ham}) can be chosen in the form
\begin{equation}\label{spin_oper}
\hat{S}_i = \frac{1}{N} \left[ \bm{\Sigma}\bm{B} - \frac{i}{m_i}\gamma_0 \gamma_5 [\bm{\Sigma}\times\bm{p}]\bm{B}\right],
\end{equation}
where
\begin{equation}
\frac{1}{N}=\frac{m_i}{\sqrt{m_i^2 \bm{B}^2 + \bm{p}^2 B^2_{\perp}}}.
\end{equation}
For the neutrino energy spectrum we obtain
\begin{equation}\label{spec}
E_i^s=\sqrt{m_i^2+p^2+{\mu_i}^2{\bf{B}}^2 +2{\mu_i}s\sqrt{m_i^2{\bf{B}}^2+p^2B_\bot^2}},
\end{equation}
where $s=\pm 1$ correspond to two different eigenvalues of the Hamiltonian (\ref{Ham}) and $p = |\bm{p}|$. Hence, we specify the neutrino spin states
as the stationary states for the Hamiltonian in the presence of the magnetic field, contrary to the customary approach to the description of neutrino oscillations when the helicity states are used. It should be noted that in case we neglect the longitudinal component of the magnetic field $B_\| =0$ the energy spectrum (\ref{spec}) coincides with the energy spectrum of a neutron  \cite{Ternov}.

The spin operator $\hat{S}_i$ commutes with the Hamiltonian $\hat{H_i}$, and for the neutrino stationary states we have
\begin{equation}
\hat{S}_i \ket{\nu_i^s} = s \ket{\nu_i^s}, s = \pm 1,
\end{equation}
and
\begin{equation}\label{normalizaton}
\braket{\nu_i^s|\nu_k^{s'}} = \delta_{ik}\delta_{ss'}.
\end{equation}
Following this line, the corresponding projector operators can be introduced
\begin{equation}\label{proj_oper}
\hat{P}^{\pm}_i = \frac{1 \pm \hat{S}_i}{2}.
\end{equation}
It is clear that projectors act on the stationary states as follows

\begin{equation}
\braket{\nu_k^{s'}|\hat{P}^{s}_i |\nu_i^{s}} = \delta_{ik}\delta_{ss'}.
\end{equation}

Now in order to solve the problem of the neutrino flavour $\nu_e^L \leftrightarrow \nu_{\mu}^L$, spin $\nu_e^L \leftrightarrow \nu_{e}^R$ and spin-flavour $\nu_e^L \leftrightarrow \nu_{\mu}^R$ oscillations in the magnetic field we expand the neutrino helicity states over the neutrino stationary states
\begin{eqnarray}\label{nu_c}
\nu_i^L(t) = c_i^+ \nu_i^+(t) + c_i^- \nu_i^-(t) ,\\
\label{nu_d}
\nu_i^R(t) = d_i^+ \nu_i^+(t) + d_i^- \nu_i^-(t),
\end{eqnarray}
where $c^{\pm}_i$ and $d^{\pm}_i$ are independent on time.

The quadratic combinations of the coefficients $c_i^{+(-)}$ and $d_i^{+(-)}$ are given by matrix elements of the projector operators (\ref{proj_oper})

\begin{eqnarray}\label{coef_1}
|c_i^{\pm}|^2 &=& \braket{\nu_i^L|\hat{P}^{\pm}_i|\nu_i^L},\\\label{coef_2}
|d_i^{\pm}|^2 &=& \braket{\nu_i^R|\hat{P}^{\pm}_i|\nu_i^R},\\\label{coef_3}
(d^{\pm}_i)^* c_i^{\pm} &=& \braket{\nu^R_i|P_i^{\pm}|\nu^L_i}.
\end{eqnarray}
Since $|c_i^{\pm}|^2$, $|d_i^{\pm}|^2$ and $(d^{\pm}_i)^* c_i^{\pm}$ are time independent, they can be determined from the initial conditions. Note that
ultrarelativistic neutrinos are produced in a weak interaction process almost as
left-handed helicity states and in this approximation helicity and chiral states are
almost indistinguishable. It means, that the spinor structure of the neutrino initial and final states is determined by
\begin{equation}
\nu^L = \frac{1}{{\sqrt{2}L^\frac{3}{2}}} \begin{pmatrix}
                              0 \\
                              -1 \\
                              0 \\
                              1
                            \end{pmatrix}, \ \ \
\nu^R = \frac{1}{\sqrt{2}L^{\frac{3}{2}}} \begin{pmatrix}
                              1 \\
                              0 \\
                              1 \\
                              0
                            \end{pmatrix},
\end{equation}
where $L$ is the normalization length. Thus, for the quadratic combinations of the coefficients we get
\begin{equation}\label{c}
|c_i^{\pm}|^2 = \frac{1}{2} \left( 1 \pm \frac{m_i B_{\parallel}}{{\sqrt{m_i^2 B^2 + p^2 B^2_{\perp}}}} \right),\end{equation}
\begin{equation}\label{d}
|d_i^{\pm}|^2 =  \frac{1}{2} \left( 1 \mp \frac{m_i B_{\parallel}}{{\sqrt{m_i^2 B^2 + p^2 B^2_{\perp}}}} \right),
\end{equation}
\begin{equation}\label{dc}
(d_i^{\pm})^* c_i^{\pm} = \mp \frac{1}{2}\frac{p(B_1 - i B_2)}{\sqrt{m^2_i B^2 + p^2 B_{\perp}^2}}.
\end{equation}

In the case $B_{\perp} = 0$ the helicity states are stationary and
$(d_i^+)^* c_i^+=(d_i^-)^* c_i^- = |c_i^-|^2 = |d_i^+|^2 = 0$, $|c_i^+|^2 = |d_i^-|^2 = 1$.

Using eqs. (\ref{nu_c}), (\ref{nu_d}) and accounting for the fact that stationary states' propagation law has the form $\nu^s_i(t) = e^{-i E^s_i t}\nu^s_i(0)$, we get that the evolution in time (space) of the relativistic neutrino flavour state $\nu^L_e$ is given by
\begin{eqnarray}\label{nu_e_L_t}\nonumber
\nu^L_e(t) = \left(c_1^+ e^{-i E_1^+ t} \nu_1^+ + c_1^- e^{-i E_1^- t} \nu_1^- \right)\cos\theta \\
+ \left(c_2^+ e^{-i E_2^+ t} \nu_2^+ + c_2^- e^{-i E_2^- t} \nu_2^- \right)\sin\theta,
\end{eqnarray}
where $\nu_i^s \equiv \nu_i^s(0)$. In exactly the same way we can write out the decomposition of the wave function of a muon neutrino.

\section{Neutrino flavour, spin and spin-flavour oscillations in a magnetic field}
\label{sec:2}

The probability of the  neutrino flavour oscillations $\nu_e^L \leftrightarrow \nu_{\mu}^L$ is given by
\begin{eqnarray}\label{prob_nu_flavour}\nonumber
P_{\nu_e^L \rightarrow \nu_{\mu}^L}(t) =
\left|\braket{\nu_{\mu}^L|\nu_e^L(t)}\right|^2  = \\ \sin^2\theta \cos^2 \theta \big| |c_2^+|^2e^{-i E^+_2 t} + |c_2^-|^2e^{-i E^-_2 t}\\ \nonumber
 - |c_1^+|^2e^{-i E^+_1 t} - |c_1^-|^2e^{-i E^-_1 t}  \big|^2.
\end{eqnarray}
Note that since the normalization condition (\ref{normalizaton}) is satisfied, we don't use the explicit form of the neutrino stationary states wave functions to calculate the oscillation probability. The dependence of the neutrino oscillation probability on the magnetic field is due to the matrix elements of the projectors (\ref{coef_1})-(\ref{coef_3}) and the energy spectrum (\ref{spec}) field dependence.

The probability of oscillations $\nu_e^L \leftrightarrow \nu_{\mu}^L$ is simplified if one accounts for the relativistic neutrino energies ($p \gg m  $) and also for realistic values of the neutrino magnetic moments and strengths of magnetic fields ($p \gg\mu B $). In this case we have
\begin{equation}\label{energy_s}
E^{s}_i \approx p + \frac{m^2_i}{2p} + \frac{\mu_i^2 B^2}{2p} + \mu_i s B_{\perp}.
\end{equation}

It is reasonable  to suppose that $\mu B << m$, then the contribution $\frac{\mu_i^2 B^2}{2p}$ can be neglected in (\ref{energy_s}).
The assumption is justified for the most astrophysical environments for which the discuss oscillation phenomena are applicable. This can be verified by the following estimations. A neutrino magnetic moment is indeed very small.  The easiest generalization of the Standard Model gives the value ~ $10^{-20}\mu_B$ for the neutrino mass $m= 0.1 \ \ eV$ (see the Eq. (\ref{mu_D})).  Other generalizations of the Standard Model can result in much bigger values for the magnetic moment, but the present laboratory constraints provide the upper limit  ~ $10^{-11}\mu_B$. A very strong magnetic field can be found in pulsars, these are the fields of the order of the critical magnetic field  $B_0={m_e}^2/e= 4.41\times 10^{13}
\ \ Gauss$. Much stronger magnetic field are believed to exist in magnetars \cite{Potekhin:2015qsa}. Using the above values it is possible to show that the assumption $\mu B << m$ is valid at least up to the magnetic fields of order $B ~ \times 10^{17}\ \ Gauss$.

In the considered case we also have
\begin{equation}
|c_i^s|^2|c_k^{s'}|^2 \approx \frac{1}{4}.
\end{equation}

The oscillation probability (\ref{prob_nu_flavour})  is given by an interplay of several oscillations with the following six characteristic frequencies
\begin{eqnarray}
E_{1}^+ - E_{1}^- &=& 2\mu_{1} B_{\perp}, \\
E_{2}^+ - E_{2}^- &=& 2\mu_{2} B_{\perp}, \\
E_2^+ - E_1^+ &=& \frac{\Delta m^2}{2p} + (\mu_2 - \mu_1) B_{\perp}, \\
E_2^- - E_1^- &=& \frac{\Delta m^2}{2p} - (\mu_2 - \mu_1) B_{\perp}, \\
E_2^+ - E_1^- &=& \frac{\Delta m^2}{2p} + (\mu_1 + \mu_2) B_{\perp}, \\
E_2^- - E_1^+ &=& \frac{\Delta m^2}{2p} - (\mu_1 + \mu_2) B_{\perp}.
\end{eqnarray}
Finally, for the probability of flavour oscillations $\nu_e^L \leftrightarrow \nu_{\mu}^L$ we get
\begin{eqnarray}\label{flavour_final}\nonumber
P_{\nu_e^L \rightarrow \nu_{\mu}^L}(t)=  \sin^2 2\theta \\\Big\{
 \cos(\mu_1 B_{\perp} t) \cos(\mu_2 B_{\perp} t)\sin ^2 \frac{\Delta m^2}{4p} t \\ \nonumber
 +\sin^2\big(\mu_{+} B_{\perp}t) \sin^2 (\mu_{-} B_{\perp} t ) \Big\},
\end{eqnarray}
where $\mu_{\pm}=\frac{1}{2}(\mu_1 \pm \mu_2)$.

From the obtained expression (\ref{flavour_final}) a new phenomenon in the neutrino flavour oscillation in a magnetic field can be seen. It follows that the neutrino flavour oscillations in general can be modified by the neutrino magnetic moment interactions with the transversal magnetic field $B_{\perp}$. In the case of zeroth magnetic moment and/or vanishing magnetic field eq.(\ref{flavour_final}) reduces to the well known probability of the flavour neutrino oscillations in vacuum.

In quite similar evaluations we also obtain probabilities of neutrino spin $\nu_e^L \leftrightarrow \nu_{e}^R$ and spin-flavour $\nu_e^L \leftrightarrow \nu_{\mu}^R$ oscillations. In particular, for of neutrino spin  oscillations $\nu_e^L \leftrightarrow \nu_{e}^R$ we get
\begin{eqnarray}\label{spin_osc}\nonumber
P_{\nu_e^L \rightarrow \nu_e^R} = \Big\{ \sin\left(\mu_{+}B_{\perp}t\right)\cos\left(\mu_{-}B_{\perp}t\right) \\
+ \cos2\theta \sin\left(\mu_{-}B_{\perp}t\right) \cos\left(\mu_{+} B_{\perp}t \right) \Big\}^2 \\ \nonumber
- \sin^2 2\theta \sin(\mu_1 B_{\perp} t) \sin(\mu_2 B_{\perp} t)\sin^2\frac{\Delta m^2}{4p} t .
\end{eqnarray}
For the probability of the neutrino spin-flavour oscillations $\nu_e^L \leftrightarrow \nu_{\mu}^R$ we get
\begin{eqnarray}\label{spin_flavour} \nonumber
P_{\nu^L_e \rightarrow \nu^R_{\mu}}(t) = \sin^2 2\theta \\\Big\{ \sin^2(\mu_{-}B_{\perp} t) \cos^2\left(\mu_{+}B_{\perp} t\right) \\ \nonumber
 +  \sin(\mu_1 B_{\perp} t) \sin(\mu_2 B_{\perp}t) \sin^2\frac{\Delta m^2}{4p}t \Big\}.
\end{eqnarray}
Similar result for the probability was obtained in  \cite{Dvornikov:2007qy} from the study of the evolution of neutrino wavefunction
in a transverse magnetic field.

For completeness,  we also calculate within our approach the neutrino survival  probability $\nu_e^L \leftrightarrow \nu_{e}^L$ and get

\begin{eqnarray}\nonumber\label{survival}
P_{\nu_e^L \rightarrow \nu_e^L}(t) = \Big\{ \cos\left(\mu_{+}B_{\perp}t\right) \cos\left(\mu_{-}B_{\perp}t\right) \\- \cos2\theta \sin\left(\mu_{+}B_{\perp}t\right) \sin\left(\mu_{-}B_{\perp}t\right) \Big\}^2 \\ \nonumber
- \sin^2 2\theta \cos(\mu_1 B_{\perp}t)\cos(\mu_2 B_{\perp}t)\sin^2\frac{\Delta m^2}{4p}t.
\end{eqnarray}
It is just straightforward that the sum of the obtained four probabilities (\ref{flavour_final}), (\ref{spin_osc}), (\ref{spin_flavour}) and (\ref{survival}) is
\begin{eqnarray}\label{summ}
P_{\nu_e^L \rightarrow \nu_{\mu}^L}+P_{\nu_e^L \rightarrow \nu_{e}^R}+P_{\nu_e^L \rightarrow \nu_{\mu}^R}+\\ \nonumber
+P_{\nu_e^L \rightarrow \nu_{e}^L}=1.
\end{eqnarray}

As an illustration of the interplay of neutrino oscillations on different frequencies,
it is interesting to find a particular realistic set of parameters (the neutrino mass square difference, energy and magnetic moment, as well as the strength of a magnetic field) which also allows one to hope for significant phenomenological consequences. Arguing so, consider as an example the neutrino flavour oscillations $\nu^L_e \rightarrow \nu_{\mu}^L$ in the transversal magnetic field $B_{\perp}$. Obviously, the stronger the magnetic field, the greater the influence it will have on the probability of the neutrino flavour oscillations. The strongest magnetic field are expected to exist in magnetars, where the strength of the field can be of the value up to $B_{\perp} = 10^{16} \ G$.

   Consider the mass square difference $\Delta m^2=7\times 10^{-5} \ eV ^2$ and the magnetic moments $\mu_1 = \mu_2=\mu \sim 10^{-20}\mu_B$ that corresponds the Standard Model prediction (\ref{mu_D})
for neutrino masses of the order $m \sim 0.1 \ eV$.
In Fig. 1 we show the probability (\ref{flavour_final}) of the neutrino flavour oscillations $\nu^L_e \rightarrow \nu_{\mu}^L$ in the transversal magnetic field for this particular choice of parameters and the neutrino energy $p = 1 \ MeV$. It is clearly seen that the amplitude of oscillations at the vacuum frequency $\omega_{vac}=\frac{\Delta m^2}{4p}$ is modulated by the magnetic field frequency $\omega_{B}=\mu B_{\perp}$.
The corresponding oscillation length is $L=1/\mu B \sim 50 \ km $. This value indeed exceeds the typical dimensions of magnetars $R_{mgt} \sim 20 - 30 \ km$ \cite{Potekhin:2015qsa}, but the effect of the oscillation amplitude modulation, as it is clearly illustrated by the Fig. 1, is still  sufficient.

A similar phenomenon of the neutrino spin and flavour oscillations modulation by the magnetic field frequency is discussed also in \cite{Kurashvili:2017zab}, where the case $\mu_{11}=\mu_{22}$ is considered.

The probability of the neutrino spin oscillations $\nu^L_e \rightarrow \nu_{e}^R$ in the transversal magnetic field $B_{\perp} = 10^{16} \ G$ for the neutrino energy $p = 1 \ MeV$, $\Delta m^2=7\times 10^{-5} \ eV ^2$ and magnetic moments $\mu_1 = \mu_2= 10^{-20}\mu_B$ is shown in Fig. 2. The probability of the neutrino spin-flavour oscillations $\nu^L_e \rightarrow \nu_{\mu}^R$ in the transversal magnetic field $B_{\perp} = 10^{16} \ G$ for the same choice of parameters is shown in Fig. 3.

\begin{figure}
\includegraphics[width=0.5\textwidth]{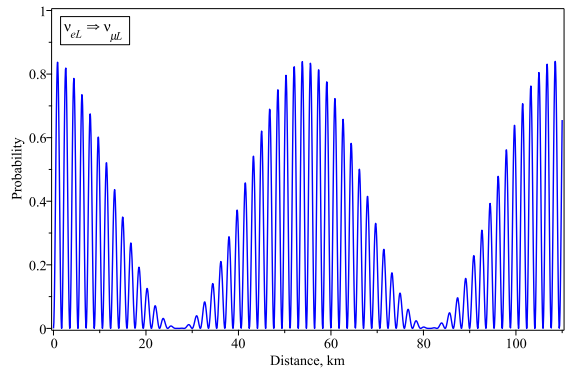}
\caption{The probability of the neutrino flavour oscillations $\nu^L_e \rightarrow \nu_{\mu}^L$ in the transversal magnetic field $B_{\perp} = 10^{16} \ G$ for the neutrino energy $p = 1 \ MeV$, $\Delta m^2=7\times 10^{-5} \ eV ^2$ and magnetic moments $\mu_1 = \mu_2= 10^{-20}\mu_B$.}
\label{fig:1}
\end{figure}

\begin{figure}
\includegraphics[width=0.5\textwidth]{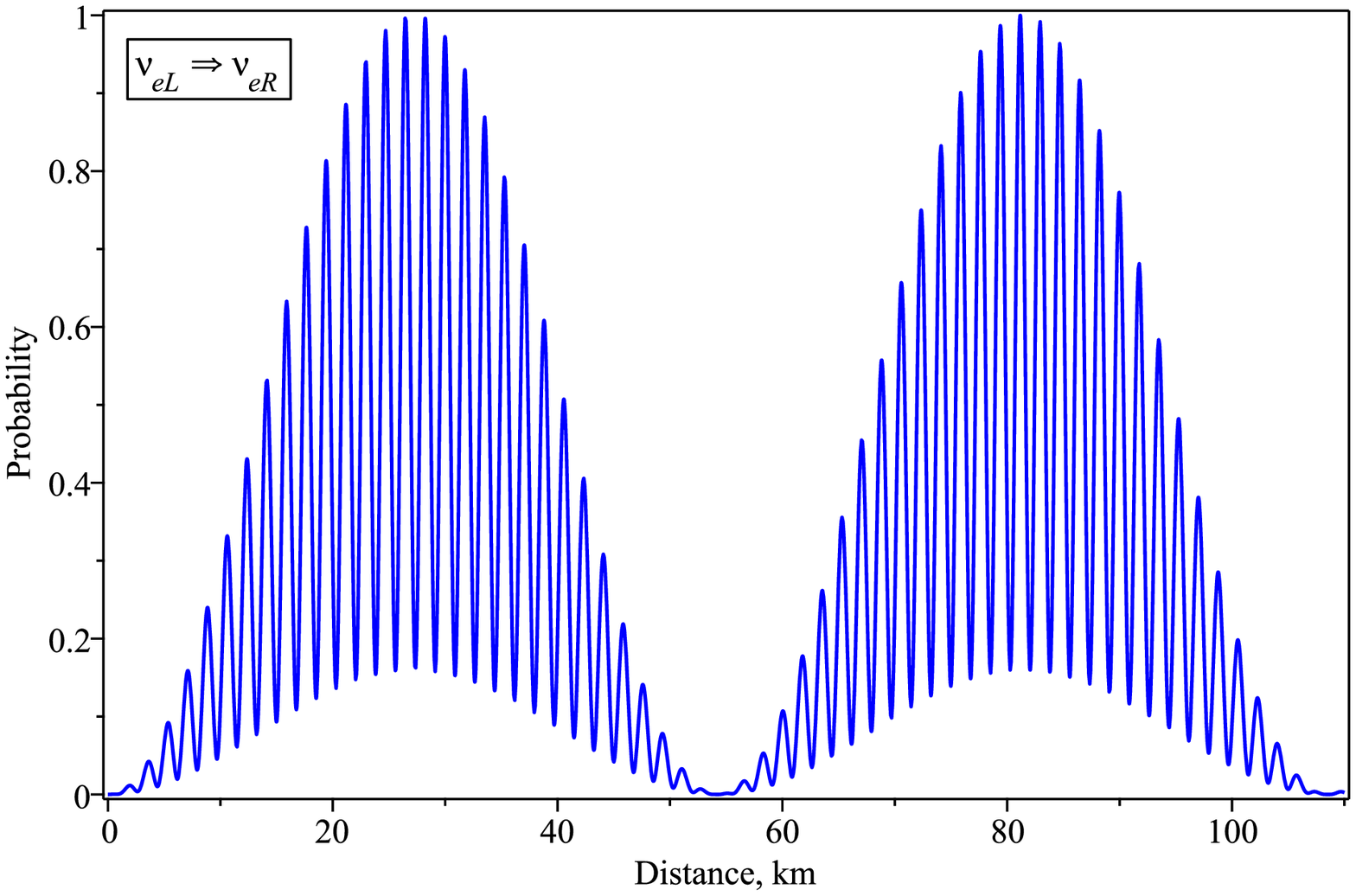}
\caption{The probability of the neutrino spin oscillations $\nu^L_e \rightarrow \nu_{e}^R$ in the transversal magnetic field $B_{\perp} = 10^{16} \ G$ for the neutrino energy $p = 1 \ MeV$, $\Delta m^2=7\times 10^{-5} \ eV ^2$ and magnetic moments $\mu_1 = \mu_2= 10^{-20}\mu_B$.}
\label{fig:2}
\end{figure}

\begin{figure}
\includegraphics[width=0.5\textwidth]{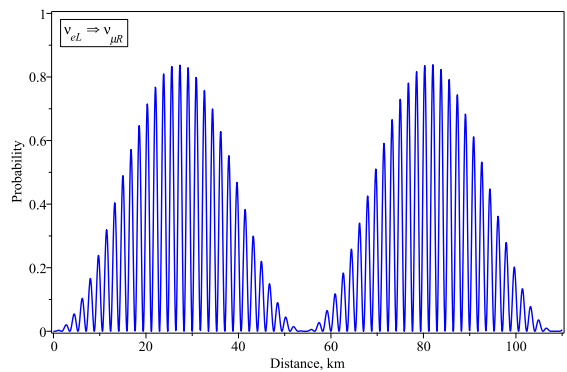}
\caption{The probability of the neutrino spin flavour oscillations $\nu^L_e \rightarrow \nu_{\mu}^R$ in the transversal magnetic field $B_{\perp} = 10^{16} \ G$ for the neutrino energy $p = 1 \ MeV$, $\Delta m^2=7\times 10^{-5} \ eV ^2$ and magnetic moments $\mu_1 = \mu_2= 10^{-20}\mu_B$.}
\label{fig:3}
\end{figure}

\section{Conclusions}
\label{concl}

We have developed a new approach to description of different types of neutrino oscillations (flavour $\nu_e^L \leftrightarrow \nu_{\mu}^L$, spin $\nu_e^L \leftrightarrow \nu_{e}^R$ and spin-flavour $\nu_e^L \leftrightarrow \nu_{\mu}^R$ oscillations) in the presence of a constant magnetic field. Our treatment of neutrino oscillations is based on the use of the exact neutrino stationary states in the magnetic field and also accounts for four neutrino states (two different mass neutrinos each in two spin states).

Consider, as an example, the probability of the neutrino spin-flavour oscillations $\nu_e^L \leftrightarrow \nu_{\mu}^R$. In literature it is often used  the probability  evaluated for the case of two neutrino species in the customary approach \cite{Akhmedov:1988uk,Lim:1987tk,Likhachev:1990ki} given by $P\sim \sin^2(\mu_{e\mu} B_{\perp}t)$ where $\mu_{e\mu} = \frac{1}{2}(\mu_2 - \mu_1)\sin2\theta$ is the transition magnetic moment in the flavour basis   \cite{Fabbricatore:2016nec,Studenikin:2017mdh}.  This probability is zero for the case $\mu_1 = \mu_2, \ \ \mu_{ij}=0, \ i\neq j$. However, the probability (\ref{spin_flavour}) of the neutrino spin-flavour oscillations $\nu_e^L \leftrightarrow \nu_{\mu}^R$ derived in our approach is not zero. In the case $\mu_1 = \mu_2 = \mu$ from (\ref{spin_flavour}) we have
\begin{equation}\label{spin_flavour_simplified}
P_{\nu_e^L \rightarrow \nu_{\mu}^R} =  \sin^2(\mu B_{\perp} t)\sin^2 2\theta \sin^2\frac{\Delta m^2}{4p}t.
\end{equation}

The neutrino spin-flavour oscillations $\nu_e^L \leftrightarrow \nu_{\mu}^R$ probability (\ref{spin_flavour}) in the particular case $\mu_1 = \mu_2$, simplified to (\ref{spin_flavour_simplified}), can be expressed as a product of two probabilities derived within the customary  two-neutrino-states approach
\begin{equation}\label{PP}
P_{\nu_e^L \rightarrow \nu_{\mu}^R} =
P_{\nu_{e}^L \rightarrow \nu_{\mu}^L}^{cust}  P_{\nu_{e}^L \rightarrow \nu_{e}^R}^{cust},
\end{equation}
where the usual expression for the neutrino spin oscillation probability
\begin{equation}\label{P_mu_B}
P_{\nu_{e}^L \rightarrow \nu_{e}^R}^{cust} = \sin^2(\mu B_{\perp}t),
\end{equation}
and the probability of the neutrino flavour oscillations
\begin{equation}\label{P_emu ll}
P_{\nu_e^L \rightarrow \nu_{\mu}^L}^{cust} = \sin^2 2\theta \sin^2\frac{\Delta m^2}{4p}t.
\end{equation}
are just the probabilities obtained in the customary approach. A
similar  neutrino spin-flavour oscillations (for the Majorana case) as a
two-step neutrino conversion processes were considered in \cite{Akhmedov:2002mf}. Since the probability of neutrino spin-flavour oscillations was supposed to be small, this effect was calculated \cite{Akhmedov:2002mf} within perturbation theory.

Now we can see that probability of spin-flavour oscillations (in the particular case $\mu_1 = \mu_2$) is a product of the customary neutrino oscillation probabilities with changing only the flavour $P_{\nu_{e}^L \rightarrow \nu_{\mu}^L}^{cust}$ and with changing only the spin state $P_{\nu_{e}^L \rightarrow \nu_{e}^R}^{cust}$.  Since in the considered case  
$$P_{\nu_{e}^L \rightarrow \nu_{e}^R}^{cust} = P_{\nu_{\mu}^L \rightarrow \nu_{\mu}^R}^{cust},$$ 
equation (\ref{spin_flavour_simplified}) can be re-written in a symmetric form:
\begin{equation}
P_{\nu_e^L \rightarrow \nu_{\mu}^R} = \frac{1}{2}\Big(P_{\nu_{e}^L \rightarrow \nu_{\mu}^L}^{cust}  P_{\nu_{\mu}^L \rightarrow \nu_{\mu}^R}^{cust} + P_{\nu_{e}^L \rightarrow \nu_{e}^R}^{cust} P_{\nu_e^R \rightarrow \nu_{\mu}^R}^{cust}\Big).
\end{equation}

In essence, this formula describes the neutrino spin-flavour oscillations probability as the sum of contributions from the two equiprobable processes: $\nu_e^L \rightarrow \nu_{\mu}^L \rightarrow \nu_{\mu}^R$ and $\nu_e^L \rightarrow \nu_{e}^R \rightarrow \nu_{\mu}^R$. Even if the transition magnetic moment in the flavour basis is vanishing, the spin-flavour change can proceed through the two step process: the flavour change and the spin flip. Thus, whereas within the customary approach the probability of spin-flavour oscillations describes just the simultaneous change of flavour and spin through the transition magnetic moment $\mu_{e\mu}$, eq. (\ref{spin_flavour_simplified}) allows spin-flavour oscillation as the sequential process. Returning to the general case when $\mu_1\neq\mu_2$, eq. (\ref{spin_flavour}) accounts for both these possibilities.

In the same way one can simplify the probability of neutrino flavour oscillations $\nu_{e}^L \rightarrow \nu_{\mu}^L$ to
\begin{eqnarray}\label{flavour_simplified} \nonumber
P_{\nu_{e}^L \rightarrow \nu_{\mu}^L} &=& \left( 1 - \sin^2(\mu B_{\perp}t) \right) \sin^2 2\theta \sin^2 \frac{\Delta m^2}{4p}t  \\ &=& \left(1 - P_{\nu_{e}^L \rightarrow \nu_{e}^R}^{cust}\right)
P_{\nu_{e}^L \rightarrow \nu_{\mu}^L}^{cust}  ,
\end{eqnarray}
The customary expression (\ref{P_emu ll}) for the neutrino flavour oscillation probability is modified by the factor $1 - P_{\nu_{e}^L \rightarrow \nu_{e}^R}^{cust}$. Since the transition magnetic moment in the flavour basis is absent in the case $\mu_1 = \mu_2$, the process $\nu_e^L \rightarrow \nu_e^R$ is the only way for spin flip, and then $1 - P_{\nu_{e}^L \rightarrow \nu_{e}^R}^{cust}$ should be interpreted as the probability of not changing the spin polarization. And consequently, this multiplier subtracts the contribution of neutrinos which changed helicity to the probability of flavour oscillations.

Similar  factor $1 - P_{\nu_{e}^{L} \rightarrow \nu_{\mu}^{L}}^{cust}$ modifies the  probability of spin oscillations $\nu_{e}^L \rightarrow \nu_{e}^R$:
\begin{eqnarray}
\label{spin_simplified} \nonumber
P_{\nu_{e}^L \rightarrow \nu_{e}^R} &=& \left[ 1 - \sin^2 2\theta \sin^2\left(\frac{\Delta m^2}{4p}t\right) \right]\sin^2(\mu B_{\perp}t) \\  &=&
\left( 1 - P_{\nu_{e}^{L} \rightarrow \nu_{\mu}^{L}}^{cust} \right)P_{\nu_{e}^L \rightarrow \nu_{e}^R}^{cust}  .
\end{eqnarray}

The neutrino survival probability $P_{\nu_e^L \rightarrow \nu_{e}^L}$ is constructed as the product of the standard probabilities of preserving neutrino flavour and preserving the spin polarization:
\begin{eqnarray}\label{survival_simplified}
P_{\nu_e^L \rightarrow \nu_{e}^L} &=& \left(1 - P_{\nu_{e}^L \rightarrow \nu_{\mu}^L}^{cust}\right)  \left(1 - P_{\nu_{e}^L \rightarrow \nu_{e}^R}^{cust}\right).
\end{eqnarray}

General formulas (\ref{flavour_final}), (\ref{spin_osc}), (\ref{spin_flavour}) and (\ref{survival}) should be interpreted in the same way. Unlike in the customary approach, oscillations of each kind are not independent. The interplay between different oscillations gives rise to interesting phenomena:

1) the amplitude modulation of the probability of flavour oscillations $\nu_e^L \rightarrow \nu_{\mu}^L$ in the transversal magnetic field with the magnetic frequency $\omega_{B}=\mu B_{\perp}$ (in the case $\mu_1 = \mu_2$) and more complicated dependence on harmonic functions with $\omega_{B}$ for $\mu_1 \neq \mu_2$;

2) the dependence of the spin oscillation probability $P_{\nu_{e}^L \rightarrow \nu_{e}^R}$ on the mass square difference $\Delta m^2$;

3) the appearance of the spin-flavour oscillations in the case $\mu_1=\mu_2$ and $\mu_{12}=0$, the transition goes through the two-step processes $\nu_e^L \rightarrow \nu_{\mu}^L \rightarrow \nu_{\mu}^R$ and $\nu_e^L \rightarrow \nu_{e}^R \rightarrow \nu_{\mu}^R$.

Finally, the obtained closed expressions (\ref{flavour_final}), (\ref{spin_osc}), (\ref{spin_flavour}) and (\ref{survival}) show that the neutrino oscillation
 $P_{\nu_e^L \rightarrow \nu_{\mu}^L}(t), \ \ P_{\nu_e^L \rightarrow \nu_e^R}(t), \ \ P_{\nu^L_e \rightarrow \nu^R_{\mu}}(t)$  and  also survival $P_{\nu_e^L \rightarrow \nu_e^L}(t)$ probabilities exhibits quiet complicated interplay of the harmonic functions that are dependent on six different frequencies (31)-(36). On this basis we predict modifications of the neutrino oscillation patterns that might provide new important phenomenological consequences in case of neutrinos propagation in extreme astrophysical environments where magnetic fields are present.

\begin{acknowledgements}
The authors are thankful to Anatoly Borisov, Alexander Grigoriev, Konstantin Kouzakov, Alexey Lokhov, Pavel Pustoshny, Konstantin Stankevich and Alexey Ternov for useful discussions. This work is supported by the Russian Basic Research Foundation grants No. 16-02-01023 and 17-52-53-133.
\end{acknowledgements}


\begin{thebibliography}{}
\bibitem{Giunti:2014ixa}
  C.Giunti and A.Studenikin,
 { Rev. Mod. Phys.}  {\bf 87}, 531 ({2015})
\bibitem{Studenikin:2018vnp}
  A.~Studenikin,
  PoS EPS {\bf -HEP2017} (2017) 137
  doi:10.22323/1.314.0137
  [arXiv:1801.08887 [hep-ph]].
\bibitem{Fujikawa:1980yx}
K.Fujikawa and R.Shrock,
Phys. Rev. Lett. {\bf 45}, 963 ({1980})
\bibitem{Shrock:1982sc}
  R.~E.~Shrock,
  Nucl.\ Phys.\ B {\bf 206} (1982) 359.
  doi:10.1016/0550-3213(82)90273-5
\bibitem{Beda:2012zz}
  A.Beda, V.Brudanin, V.Egorov, D.Medvedev  et al,
  { Adv. High Energy Phys.}  {\bf 2012}, 350150 ({2012})
\bibitem{Borexino:2017fbd}
  M.~Agostini {\it et al.} [Borexino Collaboration],
  arXiv:1707.09355 [hep-ex].
\bibitem{Raffelt:1990pj}
  G.Raffelt,
  { Phys. Rev. Lett.} {\bf 64}, 2856 ({1990})
\bibitem{Viaux:2013hca}
  N.~Viaux, M.~Catelan, P.~B.~Stetson, G.~Raffelt, J.~Redondo, A.~A.~R.~Valcarce and A.~Weiss,
  Astron.\ Astrophys.\  {\bf 558} (2013) A12
  doi:10.1051/0004-6361/201322004
  [arXiv:1308.4627 [astro-ph.SR]].
\bibitem{Arceo-Diaz:2015pva}
  S.~Arceo-Dнaz, K.-P.~Schrцder, K.~Zuber and D.~Jack,
  Astropart.\ Phys.\  {\bf 70} (2015) 1.
  doi:10.1016/j.astropartphys.2015.03.006
\bibitem{Cisneros:1970nq}
  A.Cisneros,
  Astrophys.\ Space Sci.\  {\bf 10}, 87 (1971)
\bibitem{Schechter:1981hw}
  J.Schechter and J.W.F.Valle
  { Phys. Rev.} D {\bf 24}. 1883 ({1981})
\bibitem{Okun:1986na}
  L.~Okun, M.Voloshin and M.Vysotsky,
  Sov. Phys. JETP {\bf 64}, 446 (1986)
\bibitem{Akhmedov:1988uk}
  E. Akhmedov,
 Phys. Lett. B {\bf 213}, 64 (1988)
\bibitem{Lim:1987tk}
  C.-S.Lim  and W. Marciano,
Phys. Rev. D {\bf 37}, 1368 (1988)
\bibitem{Likhachev:1990ki}
  G.~G.~Likhachev and A.~I.~Studenikin,
  J.\ Exp.\ Theor.\ Phys.\  {\bf 81} (1995) 419
   [Zh.\ Eksp.\ Teor.\ Fiz.\  {\bf 108} (1995) 769].
\bibitem{Akhmedov:1988hd}
  E.Akhmedov and M.Khlopov,
  Mod. Phys. Lett.A {bf 3}, 451 (1988)
\bibitem{Fabbricatore:2016nec}
  R.~Fabbricatore, A.~Grigoriev and A.~Studenikin,
  J.\ Phys.\ Conf.\ Ser.\  {\bf 718} (2016) no.6,  062058
  doi:10.1088/1742-6596/718/6/062058
  [arXiv:1604.01245 [hep-ph]].
\bibitem{Pulido:1999xp}
  J.~Pulido and E.~K.~Akhmedov,
  Astropart.\ Phys.\  {\bf 13} (2000) 227
  doi:10.1016/S0927-6505(99)00121-8
  [hep-ph/9907399].
\bibitem{Akhmedov:2000fj}
  E.~K.~Akhmedov and J.~Pulido,
  Phys.\ Lett.\ B {\bf 485} (2000) 178
  doi:10.1016/S0370-2693(00)00695-X
  [hep-ph/0005173].
\bibitem{Akhmedov:2002ti}
  E.~K.~Akhmedov and J.~Pulido,
  Phys.\ Lett.\ B {\bf 529} (2002) 193
  doi:10.1016/S0370-2693(02)01254-6
  [hep-ph/0201089].
\bibitem{Akhmedov:2002mf}
  E.~K.~Akhmedov and J.~Pulido,
  Phys.\ Lett.\ B {\bf 553} (2003) 7
  doi:10.1016/S0370-2693(02)03182-9
  [hep-ph/0209192].
\bibitem{Dmitriev:2015ega}
  A.~Dmitriev, R.~Fabbricatore and A.~Studenikin,
  PoS CORFU {\bf 2014}, 050 (2015)
  [arXiv:1506.05311 [hep-ph]]
\bibitem{Studenikin:2017mdh}
  A.~Studenikin,
  EPJ Web Conf.\  {\bf 125} (2016) 04018
  doi:10.1051/epjconf/201612504018
  [arXiv:1705.05944 [hep-ph]].

\bibitem{Thomson2013}
  M.Thomson,
 ``Modern Particle Physics'',
  Cambridge University Press, 2013, doi:10.1017/CBO9781139525367.
\bibitem{Potekhin:2015qsa}
  A.~Y.~Potekhin, J.~A.~Pons and D.~Page,
  Space Sci.\ Rev.\  {\bf 191} (2015) no.1-4,  239
  doi:10.1007/s11214-015-0180-9
  [arXiv:1507.06186 [astro-ph.HE]].
\bibitem{Ternov}
  I.~M.~Ternov, V.~G.~Bagrov and A.~M.~Khapaev,
  JETP {\bf 21}, 613 (1965).
\bibitem{Dvornikov:2007qy}
  M.~Dvornikov and J.~Maalampi,
  Phys.\ Lett.\ B {\bf 657} (2007) 217
  doi:10.1016/j.physletb.2007.10.019
  [hep-ph/0701209 [HEP-PH]].
\bibitem{Kurashvili:2017zab}
  P.~Kurashvili, K.~A.~Kouzakov, L.~Chotorlishvili and A.~I.~Studenikin,
  Phys.\ Rev.\ D {\bf 96} (2017) no.10,  103017
  doi:10.1103/PhysRevD.96.103017
  [arXiv:1711.04303 [hep-ph]].


\end{thebibliography}
\end{document}